\documentclass[superscriptaddress,amsfonts,amssymb,amsmath,twocolumn,floatfix,nofootinbib]{revtex4-2}
\usepackage{txfonts}
\usepackage{siunitx}
\usepackage{tikz}\usetikzlibrary[arrows]\usetikzlibrary{decorations.markings}
\usepackage{mathrsfs}
\usepackage{graphicx}
\usepackage{hyperref}
\usepackage{amsmath}
\usepackage{color}
\usepackage{xcolor}
\usepackage{numprint}
\usepackage{float}
\usepackage{ulem}
\usepackage{physics}
\usepackage{mathtools}
\usepackage{dsfont}
\usepackage{subfigure}
\usepackage{upgreek}
\usepackage{multirow}
\usepackage{mathrsfs}
\usepackage{comment}
\usepackage{bm}

\begin{document}
\title{Electromagnetic radiation from circular orbits in Schwarzschild--de~Sitter spacetime}

\author{Jo\~ao P. B. Brito}
\email{joao.brito@icen.ufpa.br} 
\affiliation{Programa de P\'os-Gradua\c{c}\~{a}o em F\'{\i}sica, Universidade 
		Federal do Par\'a, 66075-110, Bel\'em, Par\'a, Brazil}

\author{Rafael P. Bernar}
\email{rbernar@ufpa.br} 
\affiliation{Programa de P\'os-Gradua\c{c}\~{a}o em F\'{\i}sica, Universidade 
		Federal do Par\'a, 66075-110, Bel\'em, Par\'a, Brazil}

\author{Lu\'is C. B. Crispino}
\email{crispino@ufpa.br}
\affiliation{Programa de P\'os-Gradua\c{c}\~{a}o em F\'{\i}sica, Universidade 
		Federal do Par\'a, 66075-110, Bel\'em, Par\'a, Brazil}

\date{\today}
\begin{abstract}
We investigate the electromagnetic radiation emitted by a charged particle orbiting a four-dimensional Schwarzschild--de~Sitter black hole using a semiclassical approach. We calculate the probability amplitude for the charged particle to emit a photon, from which we derive the emitted power and spectral distributions. The results are compared with those obtained for scalar fields, considering both minimal and nonminimal coupling, as well as the corresponding electromagnetic case in the Schwarzschild spacetime. It is found that the total electromagnetic power is approximately twice that of the scalar field with conformal coupling for orbits with low angular velocity, similar to the Schwarzschild case, whereas for minimal coupling, the scalar power can even exceed the electromagnetic power. We also investigate the spectral distributions and possible synchrotron emission by orbits near the photon sphere. The electromagnetic case has a much broader frequency spectrum, with a non-negligible lower multipole contribution.
\end{abstract}
\maketitle

\section{Introduction}
An object moving in the gravitational field of a black hole (BH) emits radiation through different channels, including scalar, electromagnetic, and gravitational radiation. This radiation mechanism is important to both radio and gravitational wave astronomy~\cite{EHT_sombra,EHT_sombra_SgrA,ligo1_2016}, particularly in the context of binary systems with extreme mass ratios, such as a solar-mass compact object inspiraling into a (super)massive BH at the center of a galaxy. Such systems are considered promising sources of gravitational waves for space-based interferometers~\cite{LISA,LISAsite,seoane_2018}.

A simplified model for analyzing these systems involves a point particle orbiting the BH along circular geodesics. The radiation emission is computed either classically, using Green's function technique, or semiclassically, via quantum field theory (QFT) in curved spacetimes at tree level. In the semiclassical approach, the probability of the particle emitting a \textit{quantum} is used to compute the total amount of radiation. Moreover, at leading order in perturbation theory, the point particle approximation agrees well in several aspects with fully nonlinear approaches, such as simulations of BH collisions (see, e.g., Ref.~\cite{sperhake_2011,seoane_2019}).

The semiclassical approach of QFT in curved spacetimes is interesting not only because it provides a straightforward way to calculate radiation emission, but also because it offers a simpler conceptual perspective and tools for potential extensions. Using this approach, the calculation consists of determining the one-particle-emission amplitude at first order, from which the corresponding radiation is derived and found to be equivalent to that obtained through classical methods. More importantly, the study of quantum fields in fixed curved spacetimes has yielded significant insights into the quantum nature of gravity~\cite{birrell_1982,parker_2009}, including phenomena such as particle creation by dynamic spacetimes~\cite{parker_1969} and the prediction that BHs emit radiation (Hawking radiation), suggesting the possibility of their evaporation through thermal processes~\cite{hawking_1974,hawking_1975}. A closely related phenomenon in flat spacetime is the Unruh effect~\cite{unruh_1976}, which is intrinsically related to the so-called Larmor radiation~\cite{higuchi_2025}.

The scalar field serves as a simple model that captures many qualitative features similar to those of electromagnetic (vector) and gravitational (tensor) fields. 
However, it remains interesting to directly study the well-known problem of the electromagnetic radiation of a charged particle in a gravitational field, which continues to pose both conceptual~\cite{kalinov_2015, akhmedov_2025} and technical~\cite{blaga_2016,moreira_2021} challenges. 
Moreover, it is interesting to compare the different radiation channels, such as scalar and electromagnetic radiation, because although both may be treated within a unified framework, the polarization states of the electromagnetic radiation produce significantly different spectra~\cite{Ruffini1972}. Studies in the Schwarzschild spacetime have shown that a scalar particle moving around the BH along circular orbits, near the null circular orbit, emits radiation of the synchrotron type~\cite{misner_1972,misner_et_al_1972} (see also, e.g., Ref.~\cite{crispino_2000,crispino_2008}) and that the scalar spectrum is characterized by a peak at a frequency much higher than the fundamental frequency of the orbit, while the electromagnetic radiation emitted from the same orbit has a much broader frequency spectrum~\cite{breuer_1973}.

A natural extension of these investigations is to explore cosmological spacetimes, given the experimental evidence that the Universe is currently undergoing accelerated expansion~\cite{riess_1998,perlmutter_1999,Planck2020}, suggesting the presence of a nonzero (positive) cosmological constant~\cite{carroll_2001}. Moreover, de~Sitter-like solutions play a central role in both the early Universe, during the inflationary epoch, where density perturbations may have led to the formation of primordial BHs~\cite{guth_1981,Carr1975}, and in its far future, as it is plausible that our Universe is approaching a \textit{de~Sitter} phase populated by nonevaporated BHs~\cite{Perlmutter2000,Page1976}. In this context, studying BH physics in asymptotically de~Sitter solutions of general relativity becomes particularly relevant. Additionally, the analysis of radiation emitted by point particles in circular orbits serves as an important tool for investigating extreme mass ratio inspirals in de~Sitter-like spacetimes. The simplest BH solution in a de~Sitter background is the Schwarzschild--de~Sitter (SdS) BH, which describes a static, uncharged, spherically symmetric BH in a cosmological background with a positive cosmological constant~\cite{kottler_1918,stuchlik_1999}.

We investigate the photon emission due to a charged particle in circular orbits around a SdS BH. In particular, we analyze the emitted power and the spectral distribution, with special attention to ultrarelativistic orbits. We extend and compare the results with those obtained for scalar fields in Refs.~\cite{brito_2020,brito_2024_nonmini}, where both minimal and nonminimal coupling were considered. We investigate the role of the coupling of the scalar field to the curvature scalar by comparing the electromagnetic and scalar emitted power and spectral distributions in Schwarzschild and SdS spacetimes (see, e.g., Refs.~\cite{Ruffini1972,breuer_1973,castineiras_2005,bernar_2018}). Additionally, we examine the behavior of both electromagnetic and scalar spectra for orbits near the photon sphere, where synchrotron-type radiation would dominate.

The remainder of the paper is organized as follows. In Sec.~\ref{sec_elect_field}, we analyze the electromagnetic field in a modified Feynman gauge in the background of a SdS BH. We then perform the field quantization in this static spacetime using the Gupta--Bleuler prescription. In Sec.~\ref{sec_emission}, we review the motion of a charged particle in circular orbits around the BH and present the associated classical current. The one-particle emission amplitudes are then calculated, from which the radiated power and spectral distributions are obtained.
In Sec.~\ref{sec_results}, we show a selection of representative numerical results. Discussions and final remarks are given in Sec.~\ref{sec_remarks}.  
Throughout this paper, unless otherwise stated, we adopt the metric signature $(+, -, -, -)$  and units where $G=c=\hbar=1.$

\section{Quantization of the Electromagnetic field in Schwarzschild--de~Sitter spacetime}
\label{sec_elect_field}
We perform the quantization of the electromagnetic field in the static region of SdS spacetime, which corresponds to the \textit{physical} region bounded by the event and cosmological horizons. This region is described by the line element~\cite{stuchlik_1999}
\begin{equation}
\label{SdS_line_element}
ds^2 = f(r)dt^2 - \frac{dr^2}{f(r)} - r^2(d\theta^2 + \sin^2 \theta d\phi^2),
\end{equation}
where the metric function is given by
\begin{equation}
\label{f}
f(r) = 1 - \frac{2 M}{r} - \frac{\Lambda}{3}r^2,
\end{equation}
with $\Lambda$ denoting the positive cosmological constant and $M$ the BH mass. Equation~\eqref{f} has three real roots, $r_i \coloneqq \{r_-, r_h,r_c\}$, with $r_-<0<r_h<r_c$. The radial coordinates of the event and cosmological horizons are defined by $r_h$ and $r_c$, respectively. In the extremal limit $ \Lambda \to 1/9M^2 $, the two horizons approach the radius of the photon sphere, $r_0 \coloneqq 3M$.\footnote{See Ref.~\cite{gibbons_2008} for a discussion on the independence of this radial position from the cosmological constant.} At this limiting value for the cosmological constant, the physical region disappears.

The action for the electromagnetic field in a modified Feynman gauge is given by:
\begin{equation}
\label{action_0}
S = \int \mathcal{L}_{\mathrm{FG}} d^4x,
\end{equation}
where the Lagrangian density takes the form
\begin{equation}
\label{lagrangian}
\mathcal{L}_{\mathrm{FG}}=\sqrt{-g}\left(-\frac{1}{4}F^{\mu \nu}F_{\mu \nu} - \frac{1}{2}\mathfrak{G}^2\right).
\end{equation}
Here, $g\coloneqq\det( g_{\mu \nu})$, $F_{\mu\nu}=\nabla_\mu A_\nu - \nabla_\nu A_\mu$ is the electromagnetic tensor and $\mathfrak{G}$ is defined as
\begin{equation}
\label{H}
\mathfrak{G} \coloneqq \nabla^{\sigma}A_{\sigma} + K^{\sigma}A_{\sigma}.
\end{equation}
The vector $K^\sigma$ is chosen to point in the radial direction, with its only nonvanishing component given by $K^r = f'(r)$. 

The equations of motion derived from the action are
\begin{equation}
\label{Eq_of_motion}
\nabla_{\mu}F^{\mu \nu} + g^{\mu \nu} \nabla_{\mu} \mathfrak{G} - K^{\nu} \mathfrak{G} = 0,
\end{equation}
where the mode solutions with positive frequency, associated with the timelike Killing vector field $\partial_{t},$ take the form
\begin{equation}
\label{solutions}
A_{\mu}^{\lambda n; \omega \ell m} = \eta_{\mu}^{\lambda n; \omega \ell m}(r,\theta,\phi)e^{-i\omega t} \quad (\omega>0).
\end{equation}
The label $n$ distinguishes modes that are purely incoming from the past cosmological horizon $H_c^-$ ($n=\mathrm{in}$) and those incoming from the past (white hole) horizon $H^{-}$ ($n=\mathrm{up}$). In the physical region, the causal structure of SdS spacetime is similar to the Schwarzschild exterior region~\cite{castineiras_2003}.
The index $\lambda$ in Eq.~\eqref{solutions} stands for the mode polarization. The physically relevant modes are those labeled by $\lambda= I ,\, I\!I$ (the two photon polarizations), which are defined as those that satisfy both the equations of motion \eqref{Eq_of_motion} and the gauge condition $\mathfrak{G} = 0$, while not being pure gauge\footnote{The pure-gauge modes are those that are written as the gradient of a scalar function.} for $\ell \geq 1$. The physical modes are associated with physical states in the Fock space. See Refs.~\cite{higuchi_1987,crispino_2001,castineiras_2005} for further details.

In the notation $A_\mu=(A_t,A_r,A_\theta,A_\phi)$, the physical modes $A_{\mu}^{I n; \omega \ell m}$ and $A_{\mu}^{I\!I n; \omega \ell m}$ are expressed as~\cite{crispino_1998,castineiras_2005}
\begin{eqnarray}
\label{A_I}
A_{\mu}^{I n; \omega \ell m} &=& \left(0, \frac{\varphi_{\omega \ell}^{I n}}{r^2}, \frac{f(r)}{L} \frac{d \varphi_{\omega \ell}^{I n}}{dr} \partial_{\theta}, 
\frac{f(r)}{L} \frac{d\varphi_{\omega \ell}^{I n}}{dr} \partial_{\phi} \right) Y_{\ell m} e^{- i \omega t}, \\
A_{\mu}^{I\!I n; \omega \ell m} &=& \left(0,0, \varphi_{\omega \ell}^{I\!I n} Y_{\theta}^{\ell m} , \varphi_{\omega \ell}^{I\!I n} Y_{\phi}^{\ell m} \right)e^{- i \omega t},\label{A_II} 
\end{eqnarray}
where $L \coloneqq \ell(\ell+1).$ Note that the $\ell = 0$ modes do not satisfy the gauge condition, i.e., the gauge condition equation, $\mathfrak{G}=0$, cannot be solved for $\ell=0$. Hence, the physical solutions have $\ell >0$. The remaining two polarizations correspond either to pure-gauge modes (which are orthogonal to the physical modes with respect to the inner product defined below) or to nonphysical modes, i.e., modes that do not satisfy the gauge condition and possess negative norm. We note that because the charge is moving around the SdS BH along circular orbits, the two physical modes labeled $I$ and $I\!I$ couple to the charge current and contribute to photon emission. 

The radial modes $\varphi_{\omega \ell}^{\lambda n}(r)$ appearing in Eqs.~\eqref{A_I} and ~\eqref{A_II} satisfy the following differential equation:
\begin{equation}
\label{dif_equation_varphi}
f(r)\frac{d}{dr}\left(f(r)\frac{d}{dr}\varphi_{\omega \ell}^{\lambda n}(r)\right) + \left(\omega^2 - V_{\mathrm{eff}}(r)\right)\varphi_{\omega \ell}^{\lambda n}(r) = 0,
\end{equation}
with the effective potential given by
\begin{equation}
\label{potential}
V_{\mathrm{eff}}(r) = f(r)\frac{\ell(\ell+1)}{r^2}.
\end{equation}
In Eqs.~\eqref{A_I} and \eqref{A_II}, $Y_{\ell m} = Y_{\ell m}(\theta, \phi)$ and $Y^{\ell m}_{i} = Y^{\ell m}_{i}(\theta, \phi)$, with $i \in \left\{ \theta, \phi \right\}$, are the scalar and vector spherical harmonics, respectively~\cite{NIST_handbook,higuchi_1987,higuchi_1987b}. These functions are expressed as
\begin{equation}
    \label{eq_spherical}
 Y_{\ell m}(\theta, \phi) = \sqrt{\frac{2\ell + 1}{4\pi} \frac{(\ell - m)!}{(\ell + m)!}} \, P_{\ell}^{m}(\cos\theta) \, e^{i m \phi}
\end{equation}
and
\begin{equation}
    \label{eq_vec_spherical}
Y_{i}^{\ell m}(\theta, \phi) = \left[ \frac{1}{\sqrt{\ell(\ell+1)}} \right] \epsilon_{i j} \, \partial^{j} Y_{\ell m}(\theta, \phi), \quad (\ell \geq 1),
\end{equation}
where $\epsilon_{ij}$ is the totally antisymmetric tensor on the 2-sphere (with indices representing the coordinates $\theta$ and $\phi$), with the components defined by $\epsilon_{\theta\theta} = \epsilon_{\phi\phi} = 0$ and $\epsilon_{\theta\phi} = -\epsilon_{\phi\theta} = \sin\theta$.

The effective potential $V_{\text{eff}}(r)$ vanishes asymptotically near both the event horizon and the cosmological horizon. Therefore, Eq.~\eqref{dif_equation_varphi} admits approximate analytical solutions in these asymptotic regions, i.e.,
\begin{equation}
		\label{asymptotic_in}
\varphi^{\lambda \mathrm{in}}_{\omega \ell} = B_{\omega \ell}^{\lambda \mathrm{in}} \begin{cases}
 e^{- i \omega x} + \mathcal{R}^{\lambda \mathrm{in}}_{\omega \ell}e^{i \omega x}, & \quad x \to +\infty, \\
\mathcal{T}^{\lambda \mathrm{in}}_{\omega \ell}e^{- i \omega x}, & \quad x \to - \infty,
\end{cases}
		\end{equation}
		\begin{equation}
		\label{asymptotic_up}
\varphi^{\lambda \mathrm{up}}_{\omega \ell} = B^{\lambda \mathrm{up}}_{\omega \ell} \begin{cases}
e^{i \omega x} + \mathcal{R}^{\lambda \mathrm{up}}_{\omega \ell} e^{- i \omega x}, & \quad x \to - \infty, \\
 \mathcal{T}^{\lambda \mathrm{up}}_{\omega \ell} e^{i \omega x}, & \quad x \to + \infty,
\end{cases}
		\end{equation}
where $B^{\lambda n}_{\omega \ell}$ are overall normalization constants, and $\mathcal{T}^{\lambda n}_{\omega \ell}$ and $\mathcal{R}^{\lambda n}_{\omega \ell}$ are the transmission and reflection amplitudes, respectively. The tortoise coordinate $x$ is written as
\begin{equation}
    \label{tortoise}
    x(r) = \sum_{r_i} \frac{1}{f'(r_i)} \log \abs{1-\frac{r_i}{r}},
\end{equation}
which satisfies $x\to- \infty$ as $r \to r_h$ and $x\to+ \infty$ as $r \to r_c.$ From the Wronskian relations, we get $\abs{\mathcal{T}^{\lambda n}_{\omega \ell}}^2 + \abs{\mathcal{R}^{\lambda n}_{\omega \ell}}^2 = 1.$

Let $\mathcal{V}^{\mu}$ be a conserved current associated with two solutions $A^{(i)}_\mu$ and $A^{(j)}_\mu$, where $(i)$ and $(j)$ represent the collective mode labels $\lambda n; \omega \ell m.$ The current is defined as
\begin{equation}
\label{symplectic-current}
\mathcal{V}^{\mu}[A^{(i)},A^{(j)}] = i \left[\overline{A_{\sigma}^{(i)}} \Pi^{\mu \sigma}_{(j)} - \overline{\Pi^{\mu \sigma}_{(i)}}A_{\sigma}^{(j)}\right],
\end{equation}
where $\Pi_{(i)}^{\mu \nu}$ is the canonical conjugate momentum current associated with the solution $A_{\nu}^{(i)},$ given by
\begin{eqnarray}
\label{conjugate_momenta}
\Pi_{(i)}^{\mu \nu} &= & \frac{1}{\sqrt{-g}}\frac{\partial \mathcal{L}}{\partial[\nabla_{\mu}A_{\nu}]}\Big\vert_{A_{\mu} = A_{\mu}^{(i)}} \nonumber\\
& = & \left[- F^{\mu \nu} - g^{\mu \nu} \mathfrak{G} \right]\big\vert_{A_{\mu} = A_{\mu}^{(i)}}.
\end{eqnarray}

We normalize the modes using a generalized Klein--Gordon inner product, given by
\begin{equation}
\label{KG}
\left(A^{(i)},A^{(j)} \right) = \int_{\Sigma} d\Sigma\, n_{\mu} \mathcal{V}^{\mu}[A^{(i)},A^{(j)}],
\end{equation}
where $\Sigma$ is a Cauchy hypersurface in the
physical region, and $n^{\mu}$ is the future-directed unit normal to $\Sigma$.
For the physical modes with $\lambda = I,\,I\!I,$ we impose the  orthonormality condition
\begin{equation}
\label{ortogonality_relation}
\left(A^{\lambda n; \omega \ell m},A^{\lambda' n'; \omega' \ell' m'} \right) = \delta_{\lambda \lambda'}\delta_{n n'}\delta_{\ell \ell'}\delta_{m m'} \delta(\omega - \omega').
\end{equation}
 
Using Eqs.~\eqref{asymptotic_in},~\eqref{asymptotic_up},~\eqref{KG}, and~\eqref{ortogonality_relation}, we derive the overall normalization constants for the physical modes. These constants are given by
\begin{equation}
\label{overall_normalization}
\abs{B^{I n}_{\omega \ell}} = \sqrt{\frac{\ell(\ell+1)}{4\pi \omega^3}}, \quad \abs{B^{I\!I n}_{\omega \ell}} = \frac{1}{\sqrt{4\pi\omega}}.
\end{equation}

The quantum field operator $\hat{A}_{\mu}$, related to the classical field $A_\mu$, is expanded in terms of positive- and negative-frequency modes as
\begin{equation}
\label{field_operator}
\hat{A}_{\mu} = \sum_{\lambda, n, \ell, m} \int_{0}^{\infty}d\omega \left[\hat{a}_{(i)} A_{\mu}^{(i)} + \hat{a}_{(i)}^{\dagger} \overline{A_{\mu}^{(i)}} \right],
\end{equation}
where the overline denotes complex conjugation.
The annihilation and creation operators, $\hat{a}_{(i)}$ and $\hat{a}_{(i)}^{\dagger}$, related to the physical states, satisfy the following nonvanishing commutation relations,
\begin{equation}
\label{commutation}
\left[\hat{a}_{\lambda n; \omega \ell m}, \hat{a}_{\lambda' n'; \omega' \ell' m'}^{\dagger} \right] = \delta_{\lambda \lambda'}\delta_{n n'}\delta_{\ell \ell'}\delta_{m m'} \delta(\omega - \omega').
\end{equation}

The relevant particle states are constructed by acting with the creation operators on the standard vacuum state $\ket{0}$, which is defined by the annihilation condition
\begin{equation}
    \hat{a}_{\lambda n; \omega \ell m}\ket{0} = 0.
    \label{eq_Boulware}
\end{equation}
In particular, the (representative) one-particle  states are given by
\begin{equation}
\label{one_particle_state}
 \ket{\lambda n; \omega \ell m} = \hat{a}_{\lambda n; \omega \ell m}^{\dagger} \ket{0}.
\end{equation}

In the next Section, we analyze the interaction between a classical current, describing an orbiting charge around the SdS BH, and the electromagnetic field $\hat{A}_\mu.$ In particular, we calculate the one-photon-emission amplitudes, from which the emitted power and spectral distributions are derived.

\section{Photon emission and emitted power}
\label{sec_emission}
\subsection{Orbiting charged particle}
\label{geodesics}
Timelike circular orbits exist within a narrow range inside the physical region of the SdS spacetime. The radial interval supporting such orbits is given by (see, e.g., Ref.~\cite{brito_2020})
\begin{equation}
\label{circular_range}
3M < r \leq r_{\mathrm{max}},
\end{equation}
where $r_{\text{max}} = \left( 3M/\Lambda \right)^{1/3}.$
The angular velocity, $\Omega$, of a particle moving on a circular orbit of radius $r=R$ is defined by
\begin{equation}
\label{angular_velocity}
\Omega \coloneqq \frac{d \phi}{dt} = \sqrt{\frac{M}{R^3} - \frac{\Lambda}{3}},
\end{equation}
which vanishes at $R = r_{\mathrm{max}}.$
The position $r_{\mathrm{max}}$ corresponds to the point where the repulsive effect of the cosmological constant exactly balances the BH gravitational attraction.  The angular velocity of the null orbit at radial coordinate $r_0=3M$ is defined by $\Omega_0 \coloneqq \Omega (r_0)$. The function $\Omega(R)$ increases monotonically from $\Omega(r_{\text{max}})=0$ to $\Omega(r_0)$. 

We may consider the point charge $q$ as orbiting the BH along timelike circular geodesics confined in the equatorial plane ($\theta = \pi/2$ and $\dot{\theta}=0$). 
Its four-velocity is given by
\begin{equation}
\label{four_velocity}
v^{\mu} = C \left(1,0,0,\Omega \right),
\end{equation}
where the normalization factor is $C= \left[f(R) - R{}^2 \Omega^2\right]^{-1/2}$, and $\Omega$ is defined in Eq.~\eqref{angular_velocity}.
The orbiting charge is described by the following conserved and normalized current:
\begin{equation}
\label{current}
j^\mu(x) = q \frac{v^\mu}{\sqrt{-g} v^0} \delta(r-R) \delta(\theta - \pi/2)\delta(\phi - \Omega t).
\end{equation}
Here, $q$ is regarded as the coupling constant that determines the magnitude of the interaction. The quantity $j^{\mu}(x)$ is conserved in the sense that it satisfies the covariant conservation law $\nabla_{\mu}j^{\mu}=0$ and it integrates to the total charge, $\int_{\Sigma}d\Sigma\,n_\mu j^{\mu} = q$, for any Cauchy hypersurface $\Sigma$.

\subsection{One-photon-emission probability}
\label{sub_sec_one_particle_emission}
The interaction part of the action encodes the coupling of the classical current $j^\mu$ to the quantum field $\hat{A}_\mu$, and is given by
\begin{equation}
\label{int_lagrangian}
\hat{S}_{\mathrm{int}} = \int j^{\mu}(x)\hat{A}_{\mu}(x) \sqrt{-g}\,d^4x,
\end{equation}
where $j^{\mu}(x)$ is given by Eq.~\eqref{current} and the field operator, $\hat{A}_{\mu}(x)$, is given by Eq.~\eqref{field_operator}. 

From Eqs.~\eqref{A_I},~\eqref{A_II},~\eqref{current}, and~\eqref{int_lagrangian}, we anticipate that the classical current of the orbiting charge will excite both photon polarization modes, $\lambda = I$ and $\lambda = I\!I$.
In other words, the interaction action, $\hat{S}_{\text{int}}$, leads to a nonvanishing probability amplitude for the emission of a physical photon with mode $n \in \left\{\text{in}, \text{up} \right\}$, polarization $\lambda \in \left\{I, I\!I \right\}$, energy $\omega,$ and angular quantum numbers $\ell$ and $m$.

To first order in perturbation theory, the probability amplitude for this one-photon-emission process is given by
\begin{eqnarray}
\label{one_particle_emission}
\mathcal{A}^{\lambda n; \omega \ell m} &=& \bra{\lambda n; \omega \ell m} i \hat{S}_{\mathrm{int}} \ket{0} \\
\label{one_particle_emission_2}
&=& i \int j^{\mu}(x) \overline{A_{\mu}^{\lambda n; \omega \ell m}}(x) \sqrt{-g}\,d^4x.
\end{eqnarray}

As previously mentioned, since the charge is moving along circular orbits, the current $j^{\mu}$ has nonvanishing components in both the time and angular (azimuthal in our case) directions. Consequently, it couples to both physical polarizations. By contrast, if the charge were undergoing radial motion, only the $\lambda = I$ mode would be excited (see, e.g., Ref.~\cite{brito_2024_elet}). Furthermore, the emitted spectrum is discrete, as is straightforwardly seen by substituting Eqs.~\eqref{A_I},~\eqref{A_II}, and ~\eqref{current} in Eq.~\eqref{one_particle_emission_2}, which yields $\mathcal{A}^{\lambda n; \omega \ell m} \propto \delta \left( \omega - m\Omega\right)$, indicating that the photon energy is $\omega_m \coloneqq m \Omega,$ with $m\geq 1$.

The emission amplitudes for the $\lambda=I$ and $\lambda=I\!I$ polarization modes are given by,
\begin{align}
\label{amplitude_I}
    \mathcal{A}^{I n; \omega \ell m} =& q \omega_m \frac{f(R)}{\ell (\ell+1)} \frac{d \overline{\varphi_{\omega \ell}^{I n}}}{dR} \overline{Y_{\ell m} \left(\frac{\pi}{2},0\right)} \int_{- \infty}^{\infty} dt \, e^{i(\omega - m \Omega) t}, \\
    \mathcal{A}^{I\!I n; \omega \ell m} =& i q \Omega \overline{\varphi^{I\!I n}_{\omega \ell}} \overline{Y_{\phi}^{\ell m} \left(\frac{\pi}{2},0 \right)} \int_{- \infty}^{\infty} dt \, e^{i(\omega - m \Omega) t}. 
    \label{amplitude_II}
\end{align}

The physical quantity of interest, the emitted power $\mathcal{W}^{\lambda n; \omega \ell m}$, is readily obtained from Eqs.~\eqref{amplitude_I} and \eqref{amplitude_II}  (see, e.g., Refs.~\cite{castineiras_2005,crispino_2008,macedo_2012,bernar_2017,bernar_2019,brito_2020,moreira_2021,brito_2021,brito_2024_nonmini} and references therein), and it is computed as a function of the angular velocity $\Omega$. The expressions for each polarization component are given by
\begin{align}
\label{eq_power_I}
    \mathcal{W}^{I n; \omega_m \ell m} =& 2\pi q^2 \omega_m^3 \frac{f(R)^2}{\left[\ell (\ell+1)\right]^2} \abs{\frac{d \varphi_{\omega_m \ell}^{I n}}{dR}}^2 \abs{Y_{\ell m} \left(\frac{\pi}{2},0\right)}^2, \\
    \label{eq_power_II}
   \mathcal{W}^{I\!I n; \omega_m \ell m} =& 2\pi q^2 m^{-2}\omega_m^3 \abs{\varphi^{I\!I n}_{\omega_m \ell}}^2 \abs{Y_{\phi}^{\ell m} \left(\frac{\pi}{2},0 \right)}^2.
\end{align}

To evaluate these expressions as functions of the charge angular velocity $\Omega,$ we invert Eq.~\eqref{angular_velocity} to obtain $R(\Omega)$.

From Eqs.~\eqref{eq_power_I} and \eqref{eq_power_II}, we define the partial emitted power for a fixed polarization $\lambda$ and mode $n$, i.e., 
\begin{equation}
    \label{partial_power}
     \mathcal{W}^{\lambda n; \ell} \coloneqq \sum_{m=1}^{\ell}  \mathcal{W}^{\lambda n; \omega_m \ell m}.
\end{equation}
We can also define other useful quantities: Summing over modes $n$ to obtain the partial power for fixed polarization $\lambda$, $\mathcal{W}^{\lambda; \ell}$; summing over polarization $\lambda$ to obtain the partial power for fixed mode $n$, $\mathcal{W}^{n; \ell}$, which are the partial power emitted to the event ($n=\text{in}$) and cosmological ($n=\text{up}$) horizons, or summing over both $n$ and $\lambda$ to obtain the partial emitted power,
\begin{equation}
    \label{partial_power}
    \mathcal{W}^{ \ell} \coloneqq \sum_{\lambda = I, I\!I} \sum_{n = \text{in}, \text{up}}  \mathcal{W}^{\lambda n; \ell}.
\end{equation}

The total emitted power, $\mathcal{W}^{\text{Tot}}$, is obtained by summing the partial power, $  \mathcal{W}^{ \ell}$, over all multipoles $\ell \geq 1$. However, in numerical computations, the $\ell$-sum must be truncated. We therefore define the total power up to a maximum multipole order $\ell_{\text{max}}$ as
\begin{equation}
    \label{tot_power}
    \mathcal{W}^{\text{Tot}} \coloneqq \sum_{\ell=1}^{\ell_{\text{max}}}   \mathcal{W}^{ \ell}.
\end{equation}
We may also define the total power for a fixed polarization, $\mathcal{W}^{\lambda;\text{Tot}}$, fixed mode, $\mathcal{W}^{n;\text{Tot}}$, or for both, $\mathcal{W}^{\lambda n;\text{Tot}}$.

It is important to note that $Y_{\ell m}(\pi/2, 0)$ \textbf{(}$ Y^{\ell m}_{\phi}(\pi/2, 0)$\textbf{)} vanishes for $\ell + m$ odd (even). Consequently, the corresponding emitted power
 $\mathcal{W}^{I n; \omega_m \ell m}$ ($\mathcal{W}^{II n; \omega_m \ell m}$) also vanishes due to the angular dependence of the emission amplitudes. Additionally, for fixed $\ell$, the dominant contributions occur for the largest allowed values of $m$: namely $m=\ell$, for $\lambda=I$ modes, and $m=\ell-1$, for $\lambda=I\!I$ modes. 
 For scalar radiation, the dominant contributions come from the modes with $m=\ell$, since the emission amplitude is also proportional to $Y_{\ell m}(\pi/2, 0)$. This dominance is caused by exponential damping factors involving $\ell-m$ or $\ell-m-1$ in the corresponding total power expressions~\cite{breuer_1973}. 
 
 The spectral distribution of the emitted radiation is derived from Eqs.~\eqref{eq_power_I} and \eqref{eq_power_II}, expressed as a function of the discrete frequencies $\omega_m = m\Omega$, with fixed orbital frequency, $\Omega$. We denote the partial spectrum as $\mathcal{W}^{\lambda n; \Omega m}$, where only the dominant term in the sum over $\ell$ is considered, i.e., $\ell=m$ for $\lambda=I$ (and the scalar spectra) and $\ell=m+1$ for $\lambda = I\!I$. 
 Our focus is on the \textit{total spectrum}, $\mathcal{W}^{\Omega m}$, which includes the contributions from both physical polarizations ($\lambda = I , I\!I$) and modes $n= \text{in}, \text{up}$. Thus, it is defined by
 \begin{equation}
     \label{total_spec}
     \mathcal{W}^{\Omega m} \coloneqq \sum_{\lambda = I, I\!I} \sum_{n = \text{in}, \text{up}} \mathcal{W}^{\lambda n; \Omega m}.
 \end{equation}
We can likewise define partial spectra keeping $\lambda$ or $n$ fixed, in analogy with the partial power discussed earlier. It is also useful to define the total spectrum \textit{normalized} by its maximum value at $m = m_0$, i.e., 
\begin{equation}
    \label{normalized_spec}
    \widetilde{\mathcal{W}}^{\Omega m} \coloneqq \frac{\mathcal{W}^{\Omega m}}{\mathcal{W}^{\Omega m_0}_{\text{max}}},
\end{equation}
where $\mathcal{W}^{\Omega m_0}_{\text{max}}$ denotes the maximum value of $\mathcal{W}^{\Omega m}$ at $m=m_0$.

To evaluate the Eqs.~\eqref{eq_power_I} and \eqref{eq_power_II}, and hence \eqref{total_spec}, we need the radial modes, $\varphi_{\omega_m \ell}^{\lambda n}$, which are obtained numerically by integrating the homogeneous differential equation, given by~\eqref{dif_equation_varphi}, subject to the boundary conditions given by Eqs.~\eqref{asymptotic_in} and \eqref{asymptotic_up}. The numerical integration is carried out over the domain $r \in \left[r_h + \epsilon, r_c - \epsilon \right]$, where $\epsilon \coloneqq 10^{-5} M$ defines a small buffer near the event and cosmological horizons to avoid numerical instabilities. The chosen parameters have been verified to yield convergent results, as presented in the next Section.

\section{Results}
\label{sec_results}
We compute the electromagnetic emitted power and compare it with that associated with scalar fields minimally and nonminimally coupled to the curvature scalar.  The magnitude of the coupling is quantified by the parameter $\xi$, where, in particular, minimal coupling corresponds to $\xi=0$, and conformal coupling corresponds to $\xi=1/6$. Our analysis focuses on these two cases, for which the emitted power has been studied in Refs.~\cite{brito_2020, brito_2024_nonmini}. Here, we extend this study to include the corresponding spectral distributions.
For convenience, we adopt a unified notation for the electromagnetic and scalar emitted power and spectral distributions, where the specific nature of the field is sometimes indicated, when necessary, by a subscript, e.g., $\mathcal{W}^{\text{Tot}}_{\text{\tiny Electromagnetic}}$ or $\mathcal{W}^{\text{Tot}}_{\text{ \tiny Scalar,} \,{\tiny \xi}}$. Thus, we also refer to the formulas introduced for the electromagnetic case when discussing the corresponding scalar quantities.
Moreover, we consider a representative value for the cosmological constant ($M^2\Lambda = 15^{-1}$), and compare it with the Schwarzschild case ($\Lambda = 0$)~\cite{Ruffini1972,chitre_1972,breuer_1973,bernar_2018}. Additionally, we extend the previous studies in Schwarzschild spacetime by (i) investigating the electromagnetic emitted power for unstable orbits; (ii) investigating the electromagnetic spectral distributions. The emitted power is presented as a function of $\Omega$, while the spectral distribution is presented as a function of the azimuthal quantum number, $m = \omega_m / \Omega$, instead of the emitted frequency $\omega_m$. The emitted power is plotted up to the maximum value of $\Omega$, which corresponds to the angular velocity of the null orbit, $\Omega_0 = \Omega(3M)$ [see Eq.~\eqref{angular_velocity}].

\subsection{The emitted power}
\label{sec_results_power}

\begin{figure*}
\includegraphics[width=.5\linewidth]{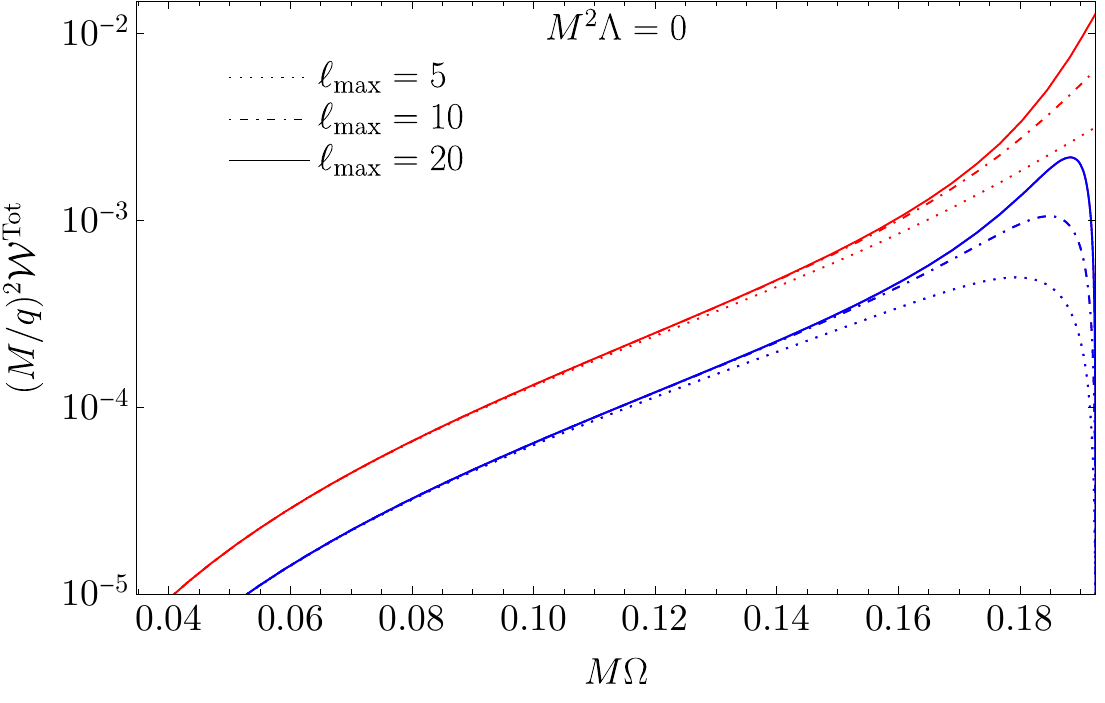}%
\includegraphics[width=.5\linewidth]{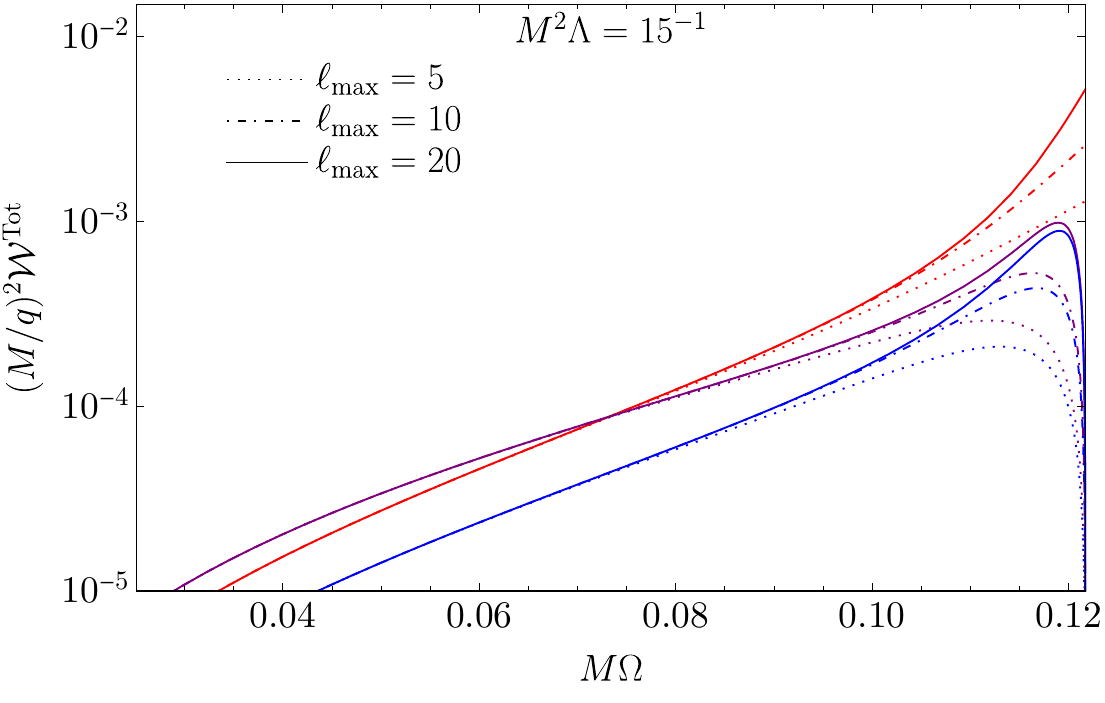}
    \caption{Total emitted power, given by Eq.~\eqref{tot_power}, for electromagnetic (red), minimally coupled scalar (purple), and conformally coupled scalar (blue) fields as a function of $M\Omega$, for $M^2 \Lambda = 0$ (left) 
    and $M^2 \Lambda = 15^{-1}$ (right). 
    The results are presented for three values of the maximum multipole number, $\ell_{\text{max}}=5,10,20$.}
\label{fig_Tot_power_comparison_Lamb_50}
\end{figure*}

\begin{figure}
\includegraphics[width=1.\linewidth]{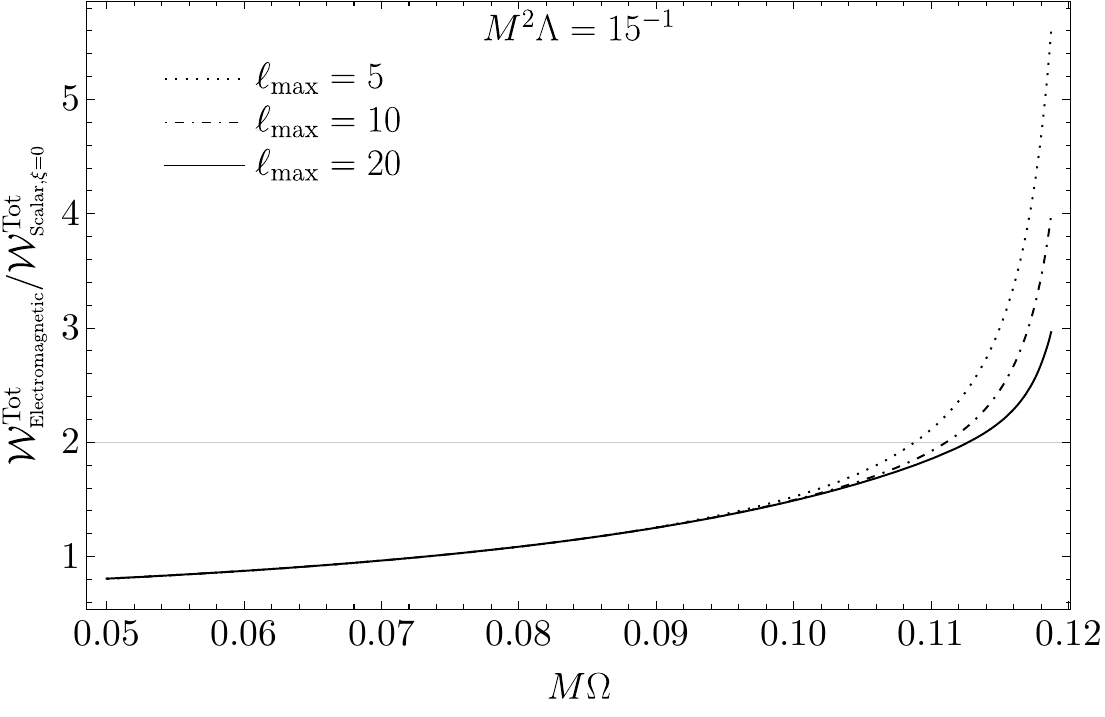}
    \caption{Ratio of the total emitted power of the electromagnetic field to that of the minimally coupled scalar field in SdS spacetime with $M^2 \Lambda = 15^{-1}$.}
    \label{fig_Tot_Power_Electromag_over_Scalar_xi0_Lamb_15}
\end{figure}

\begin{figure}
\includegraphics[width=1.\linewidth]{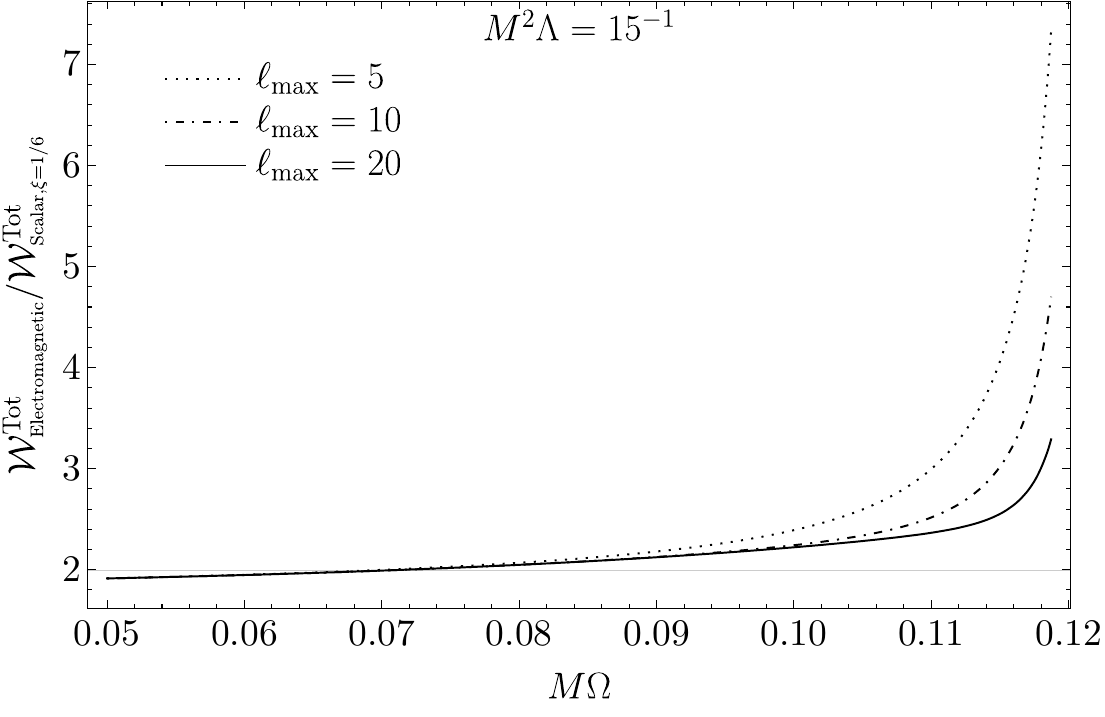}
    \caption{Ratio of the total emitted power of the electromagnetic field to that of the conformally coupled scalar field in SdS spacetime with $M^2 \Lambda = 15^{-1}$.}
\label{fig_Tot_Power_Electromag_over_Scalar_xi6_Lamb_15}
\end{figure}

\begin{figure}
\includegraphics[width=1.\linewidth]{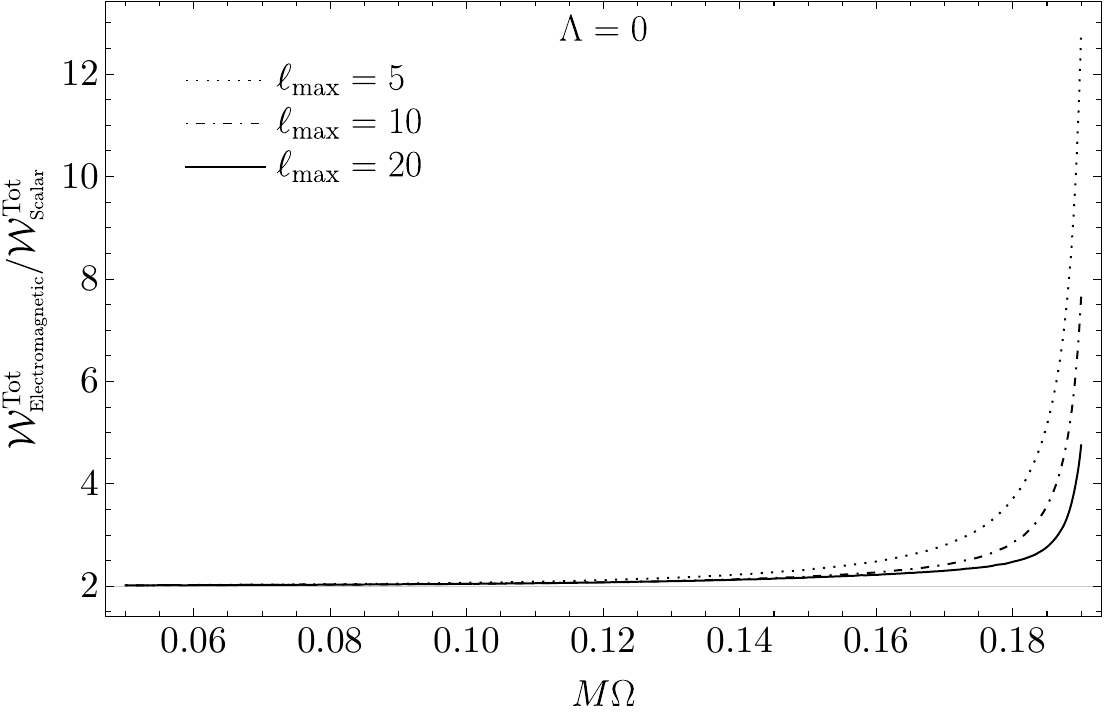}
    \caption{Ratio of the total emitted power of the electromagnetic field to that of the scalar field in Schwarzschild spacetime.}
    \label{fig_Tot_Power_Electromag_over_Scalar_xi0}
\end{figure}

Figure~\ref{fig_Tot_power_comparison_Lamb_50} shows the total electromagnetic and scalar emitted power in Schwarzschild and SdS spacetimes, given by Eq.~\eqref{tot_power}, in which the scalar field is either minimally coupled ($\xi=0$) or conformally coupled ($\xi=1/6$) to the curvature scalar. The plots are shown for three choices of $\ell_{\text{max}}.$
Interestingly, at low angular velocities, the total power emitted by the minimally coupled scalar field may exceed that of the electromagnetic field in SdS spacetime, as can be seen in the right panel. This is more clearly evidenced in Fig.~\ref{fig_Tot_Power_Electromag_over_Scalar_xi0_Lamb_15}, which shows the ratio of the electromagnetic power to the scalar power with minimal coupling. The ratio is less than unity for $M\Omega \lessapprox 0.07.$ (We note that this value decreases with decreasing $\Lambda$.)
In contrast, the electromagnetic power tends to a value slightly below twice the total power of the conformally coupled scalar field for low angular velocities, which can be seen in Fig.~\ref{fig_Tot_Power_Electromag_over_Scalar_xi6_Lamb_15}. This is consistent with the behavior observed in Schwarzschild and flat spacetimes, where the electromagnetic power approaches twice the scalar\footnote{We note that, for spacetimes with $R=0$, the minimally and conformally coupled scalar fields obey the same equations of motion.} power~\cite{castineiras_2005} (see Fig.~\ref{fig_Tot_Power_Electromag_over_Scalar_xi0}). For orbits close to the photon sphere (high angular velocities) in SdS spacetime, the behavior is similar to what happens in the Schwarzschild spacetime case, where the electromagnetic power greatly exceeds the scalar power, as can be seen in Figs.~\ref{fig_Tot_power_comparison_Lamb_50}--\ref{fig_Tot_Power_Electromag_over_Scalar_xi0}.

\begin{figure}
\includegraphics[width=1.\linewidth]{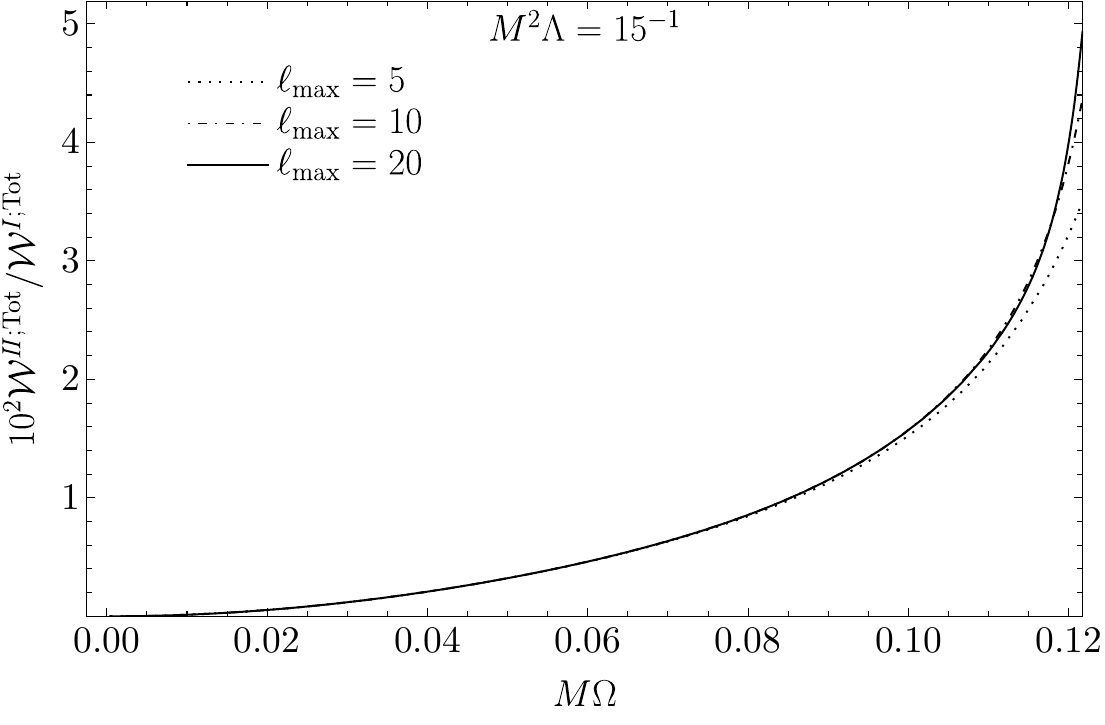}
    \caption{Relative contributions of the $\lambda= I$ and $\lambda=I\!I$ photon polarizations to the total emitted power.}
    \label{Tot_ratio_I_II_Lamb_15}
\end{figure}

It is also important to note that the two photon polarizations contribute quite differently to the emitted power, as in Schwarzschild spacetime~\cite{castineiras_2005}. For $ M^{2}\Lambda = 15^{-1}$, the contribution of the $\lambda = I\!I$ polarization to the total emitted power is negligible for orbits near $r_{\text{max}}$. Nonetheless, this contribution increases monotonically as the orbital radius decreases, reaching up to approximately $5\%$ of the $\lambda = I$ component near the photon sphere at $R = r_0$, as is shown in Fig.~\ref{Tot_ratio_I_II_Lamb_15}. 

\begin{figure}
\includegraphics[width=1. \linewidth]{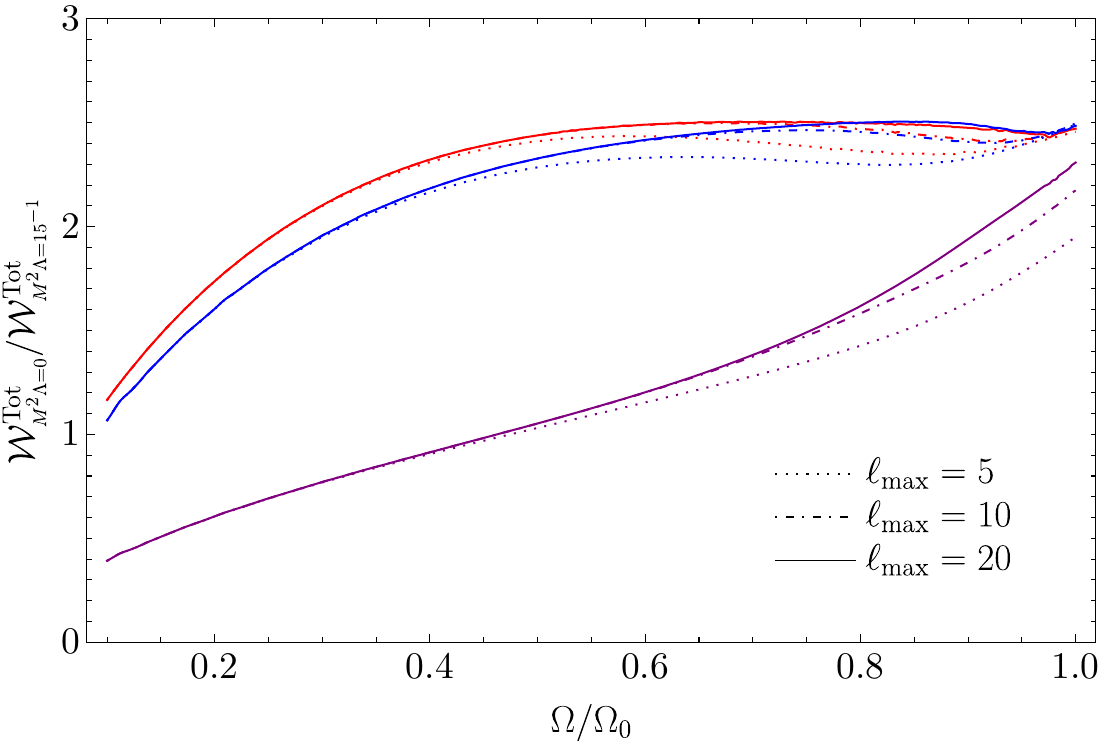}
    \caption{Ratio of the total emitted power in Schwarzschild spacetime to that in SdS spacetime with $M^2\Lambda = 15^{-1}$, for electromagnetic (red), minimally coupled scalar (purple), and conformally coupled scalar (blue) fields, plotted as a function of $\Omega/\Omega_0$, where $\Omega_0 = \Omega(3M)$. The plots are presented for three values of the maximum multipole number, $\ell_{\text{max}}=5,10,20$.}
\label{ratio_Power_Schw_SdS}
\end{figure}

Figure~\ref{ratio_Power_Schw_SdS} shows the ratio of the total emitted power in Schwarzschild spacetime to that in SdS spacetime with $M^2\Lambda = 15^{-1}$ for each field, plotted as a function of $\Omega/\Omega_0$, where $\Omega_0 = \Omega(3M)$. As we can see, the electromagnetic and scalar field ratios exhibit similar behavior if conformal coupling is considered: they increase with $\Omega/\Omega_0$, approaching approximately a constant value for $\Omega/\Omega_0 \gtrsim
 0.5$. In contrast, the ratio of the scalar emitted power in Schwarzschild spacetime to that in SdS spacetime with minimal coupling is much smaller and increases monotonically with $\Omega/\Omega_0$, due to the enhanced contribution from lower multipoles to the power in the SdS spacetime. For larger values of $\ell_{\text{max}}$, this ratio approaches those of the electromagnetic and conformal cases near the photon sphere, since in this region the emission is dominated by higher multipoles, while for low angular velocity the contribution from higher multipoles is negligible.

\begin{figure*}
\includegraphics[width=.5\linewidth]{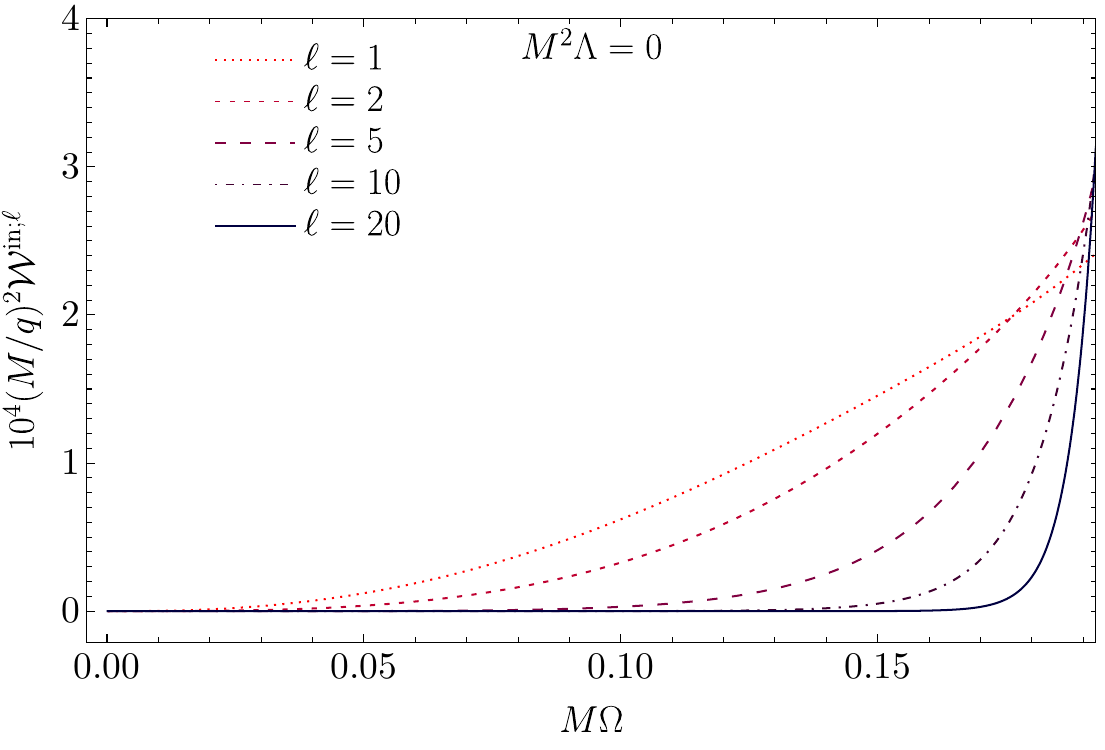}%
\includegraphics[width=.5\linewidth]{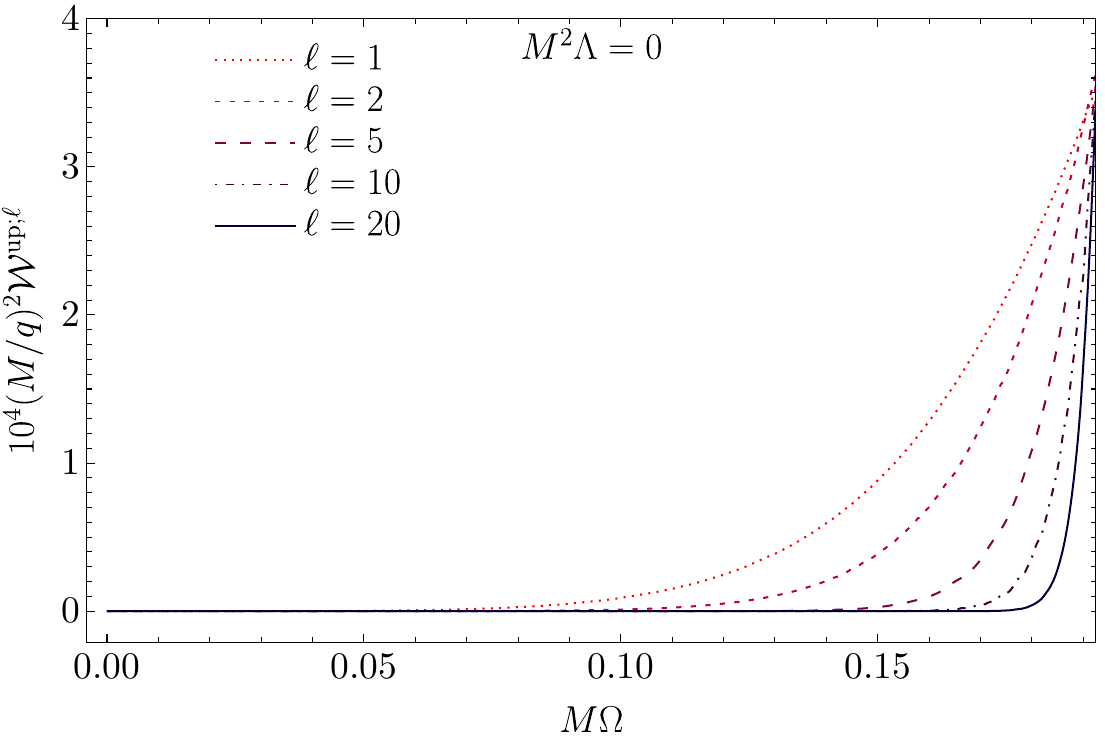}\\%
\includegraphics[width=.5\linewidth]{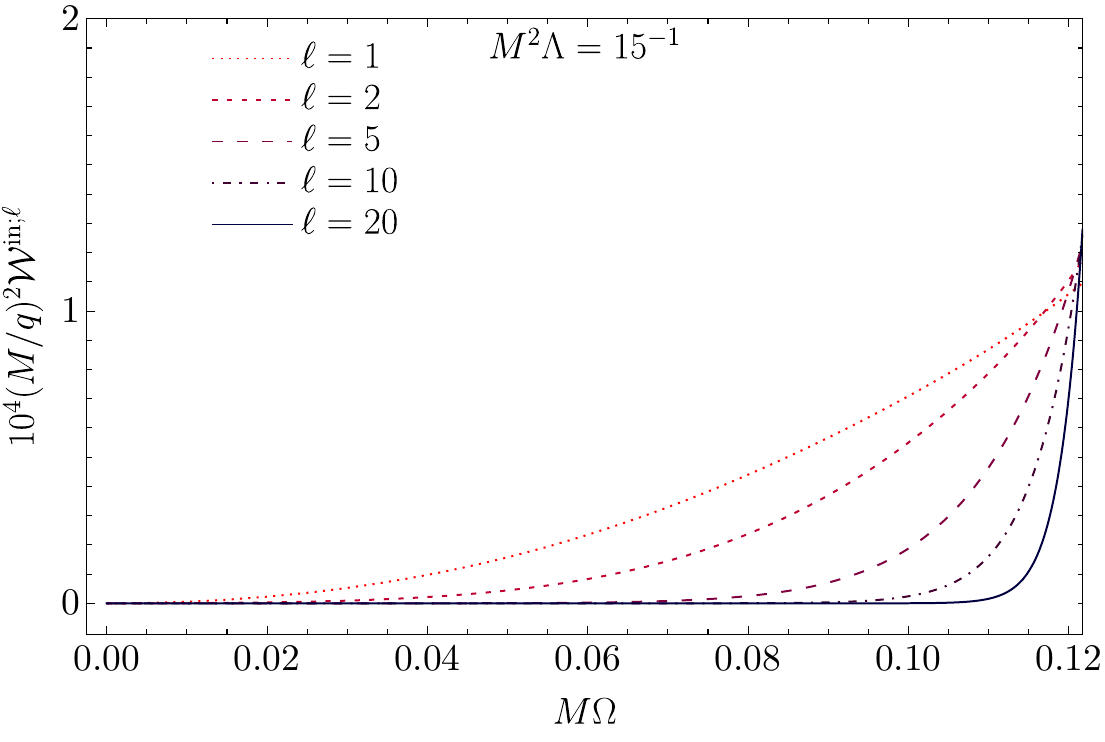}%
\includegraphics[width=.5\linewidth]{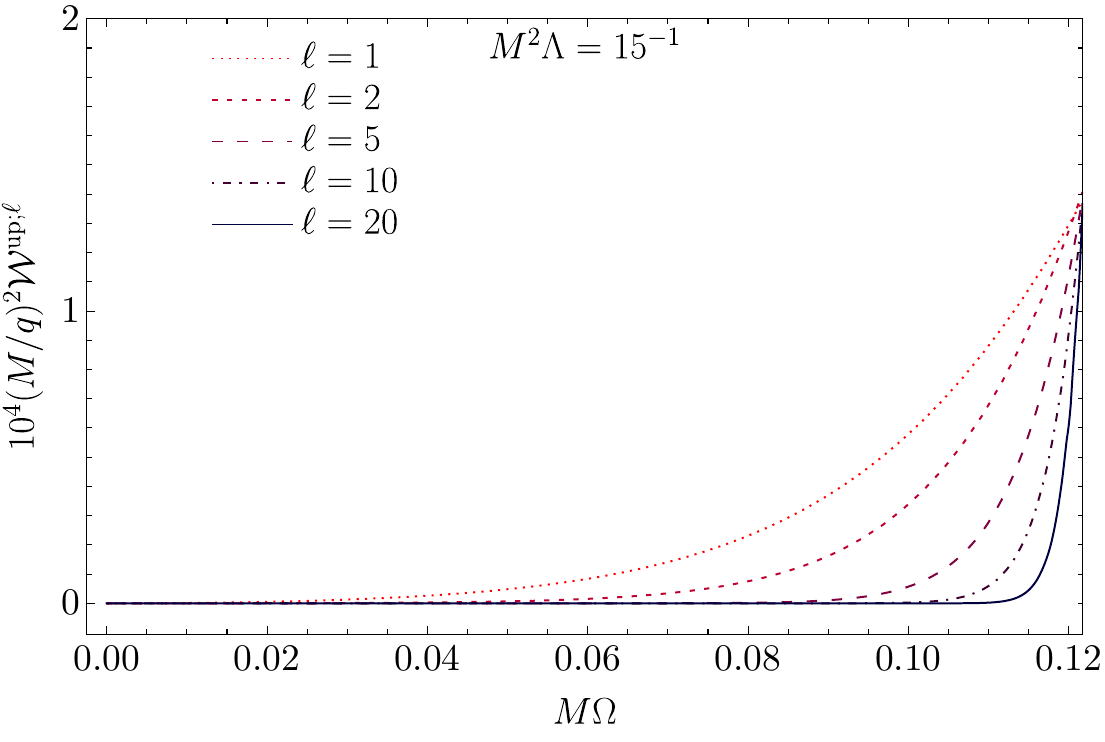}
    \caption{Electromagnetic partial emitted power for each mode $n$ in Schwarzschild spacetime (top) and SdS spacetime with $M^2 \Lambda = 15^{-1}$ (bottom), as a function of $M\Omega$. The in-mode contributions are shown on the left and the up-mode contributions on the right.}
    \label{fig_Parc_power_elet}
\end{figure*}

The radiation due to the orbiting charge is emitted either outward to the cosmological horizon or downward to the event horizon. In our framework, these emissions correspond to the in and up modes, respectively, since the charge orbit is stationary.\footnote{Emissions toward the cosmological and event horizons are more naturally associated with the so-called out and down modes, respectively, which are related to the in and up modes by complex conjugation. See, e.g. Ref.~\cite{frolov_1998}.} In Schwarzschild and SdS spacetimes, the portion of the total electromagnetic power that is emitted to the cosmological horizon exceeds the portion that is emitted to the BH. However, this behavior reverses for orbits very close to the photon sphere, where the power emitted to the event horizon becomes dominant.  This behavior can be seen in Fig.~\ref{fig_Parc_power_elet}, which shows the separate contributions of the in and up modes to the electromagnetic emitted power (we sum over polarizations $\lambda$, keeping $n$ and $\ell$ fixed) for selected values of the multipole number, $\ell$. 

Next, we analyze the electromagnetic and scalar spectra due to an orbiting particle near the photon sphere and at low angular velocities.

\subsection{The spectral distribution}

\begin{figure*}
\includegraphics[width=.5 \linewidth]{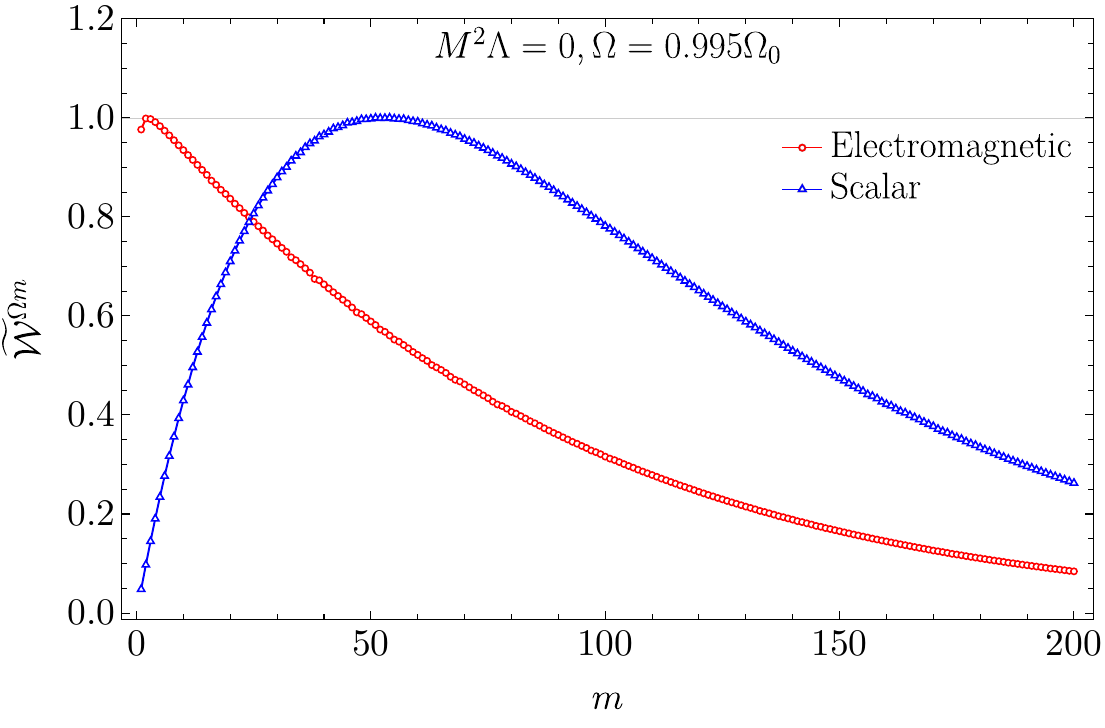}%
\includegraphics[width=.5 \linewidth]{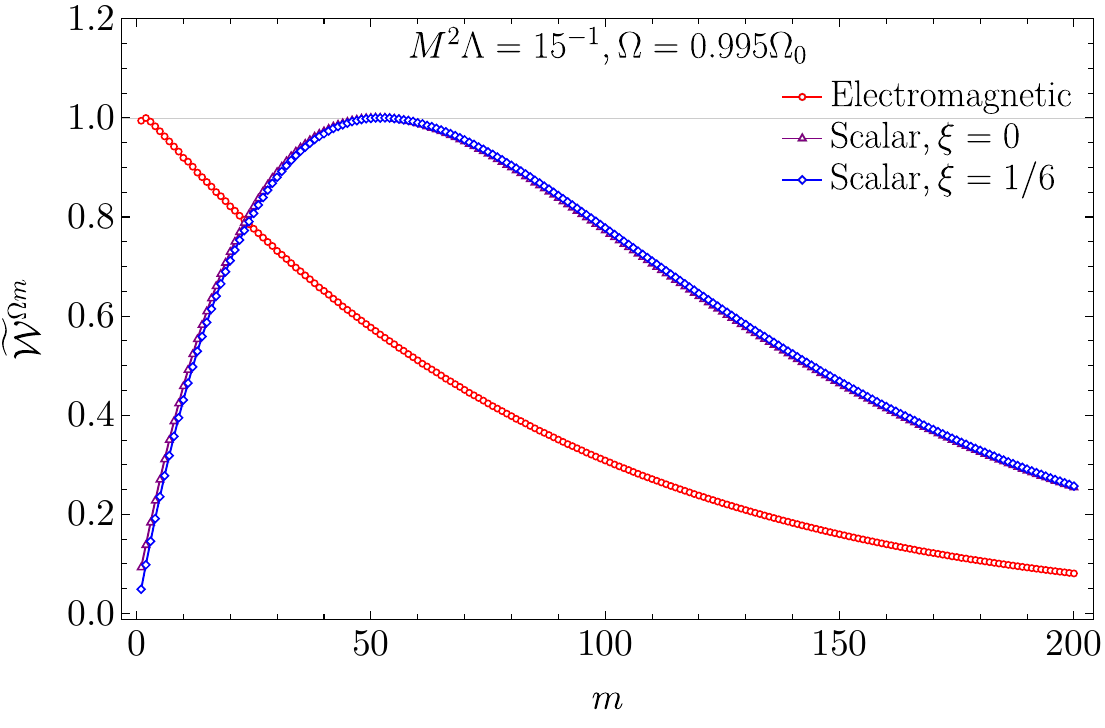}\\%
\includegraphics[width=.5 \linewidth]{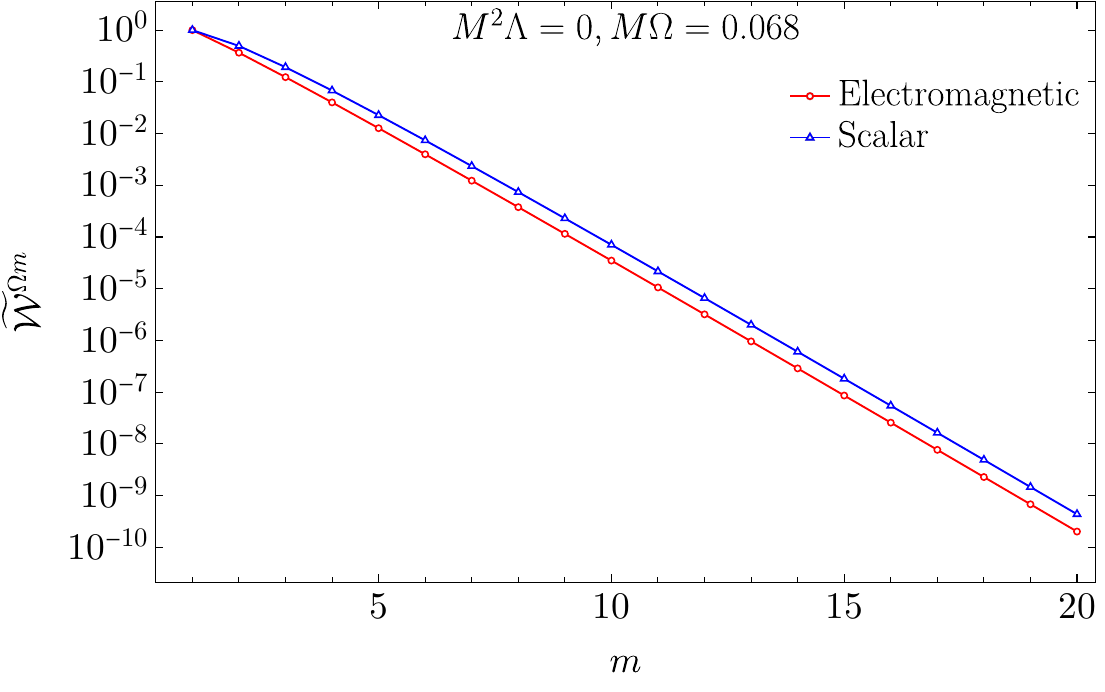}%
\includegraphics[width=.5 \linewidth]{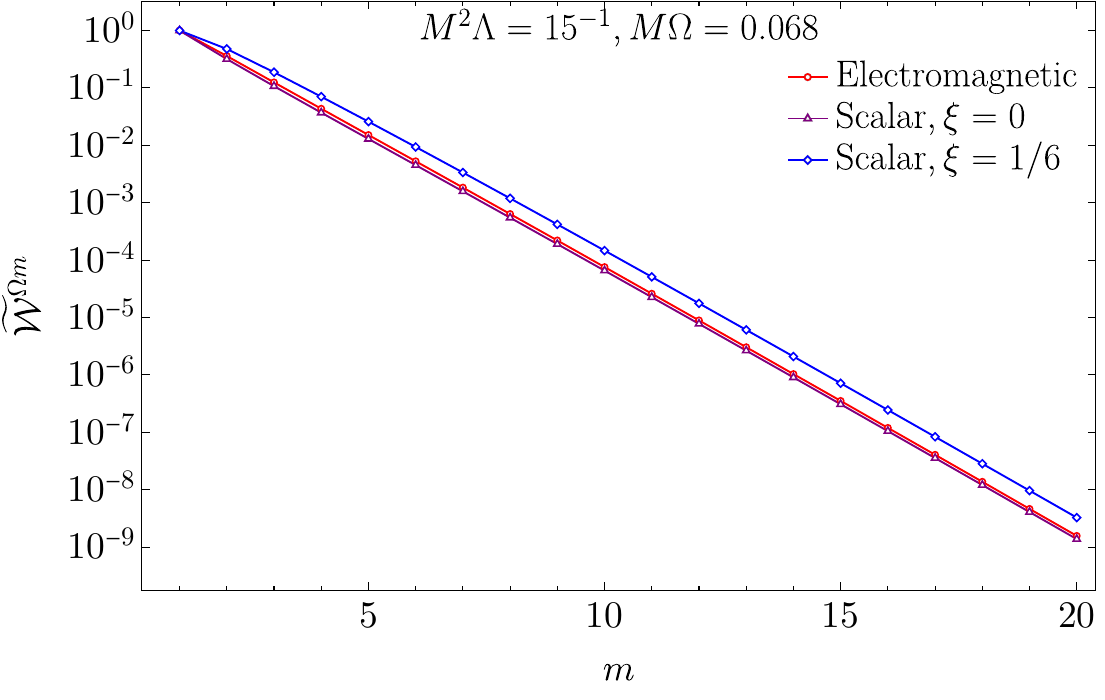}%
    \caption{Normalized spectral distributions for electromagnetic (red), minimally coupled scalar (purple), and conformally coupled scalar (blue) fields in Schwarzschild spacetime and SdS spacetime with $M^2 \Lambda = 15^{-1}$, for orbits with $\Omega = 0.995\Omega_0$ (top) and $M\Omega = 0.068$ (bottom).}
    \label{fig_spectrum_comparison}
\end{figure*}

\begin{figure*}
\includegraphics[width=.5 \linewidth]{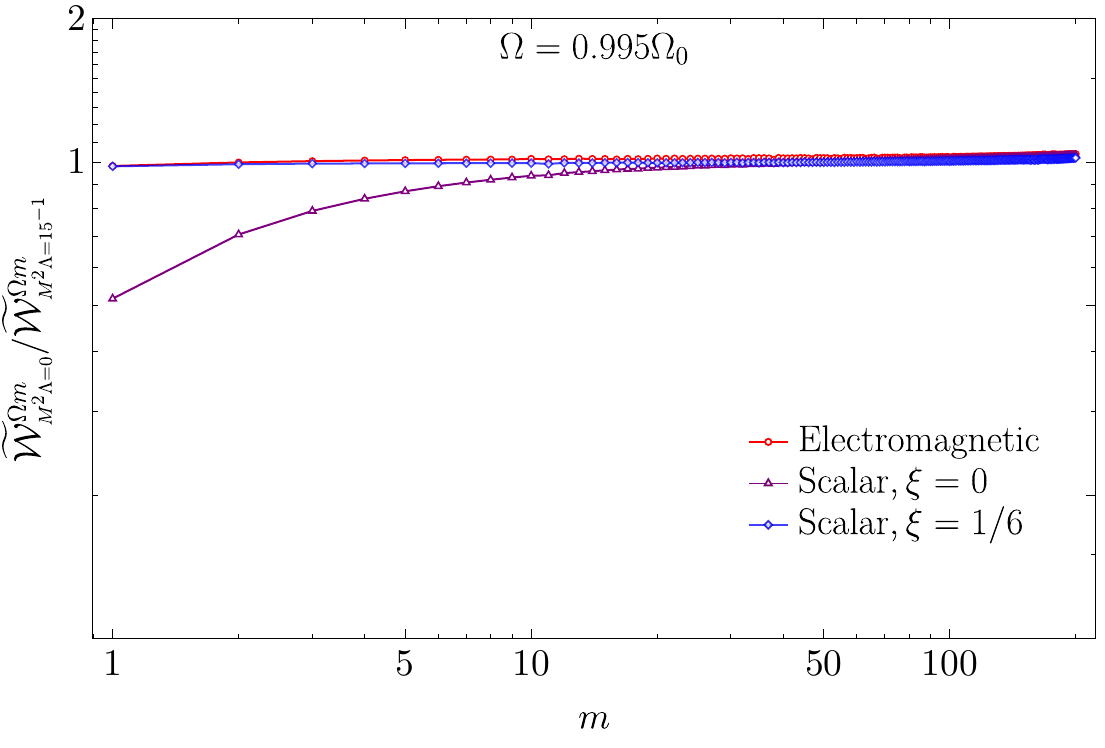}%
\includegraphics[width=.5 \linewidth]{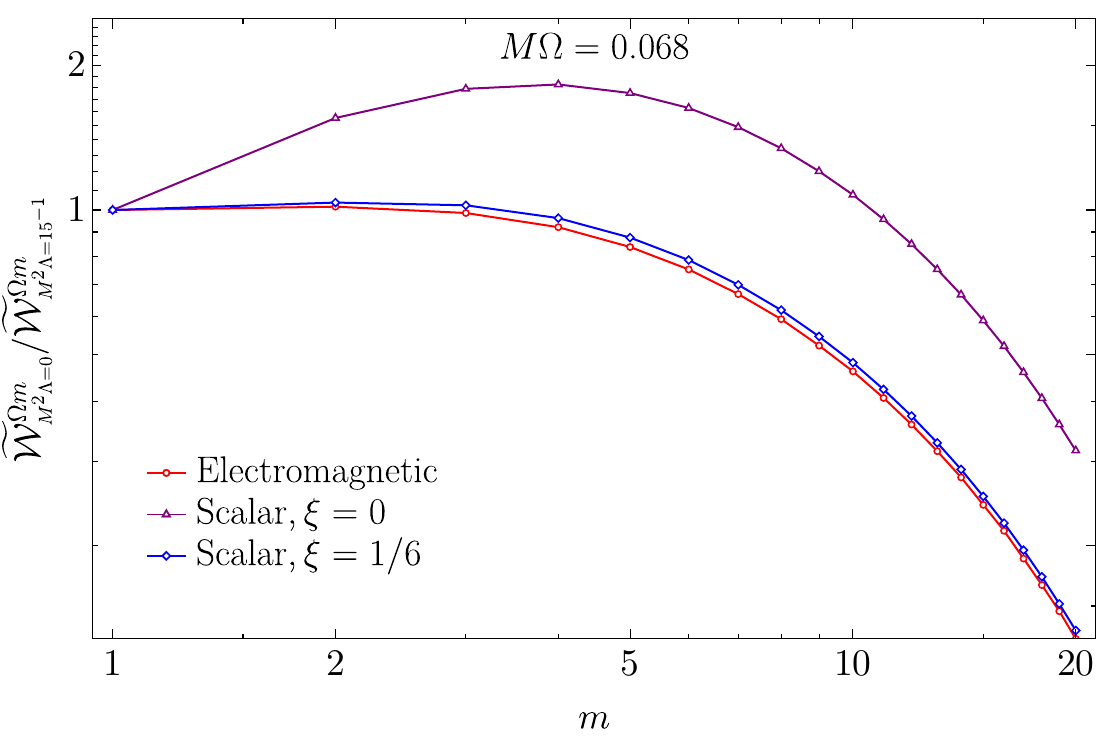}
    \caption{Ratio of the normalized spectra in Schwarzschild spacetime to that in SdS spacetime with $M^2\Lambda = 15^{-1}$, for orbits with $\Omega = 0.995\Omega_0$ (right) and $M\Omega = 0.068$ (left).}
    \label{norm_spec_ratio_Schw_SdS}
\end{figure*}

Figure~\ref{fig_spectrum_comparison} shows the normalized electromagnetic and scalar spectral distributions
in Schwarzschild and SdS spacetimes, for orbits  with $\Omega = 0.995 \Omega_0$ (near the photon sphere) and $M \Omega = 0.068$. Recall that $\Omega_0$ is the angular velocity of the null circular orbit of the corresponding spacetime. The radial position of the orbit near the photon sphere is $R\approx r_0+10^{-2} M$ in Schwarzschild spacetime and $R\approx r_0+4\times10^{-3} M$ in SdS spacetime with $ M^{2}\Lambda = 15^{-1}$. The radial position of the orbit with $M \Omega = 0.068$ corresponds to the innermost stable circular orbit in Schwarzschild spacetime, $r_{\text{isco}}=6M$, and to $R\approx 3.34 M$ in SdS spacetime.
 As can be seen, the electromagnetic and scalar spectra in Schwarzschild spacetime are similar to the electromagnetic and conformally coupled scalar spectra in SdS spacetime. This becomes more evident in the ratio of the normalized spectra in Schwarzschild spacetime to the corresponding normalized spectra in SdS spacetime, as can be seen in Fig.~\ref{norm_spec_ratio_Schw_SdS}. 
 For both electromagnetic and scalar fields with conformal coupling, the ratios are similar. For the ultrarelativistic orbit ($\Omega=0.995 \Omega_0$), it remains nearly unity across all multipoles.
 In contrast, the ratio associated with minimal coupling differs significantly from that of the electromagnetic case, particularly at lower multipoles. As discussed earlier, this is because lower multipoles are enhanced in SdS spacetime when the scalar field is minimally coupled~\cite{brito_2020}.
 
\begin{figure*}
\includegraphics[width=0.5\linewidth]{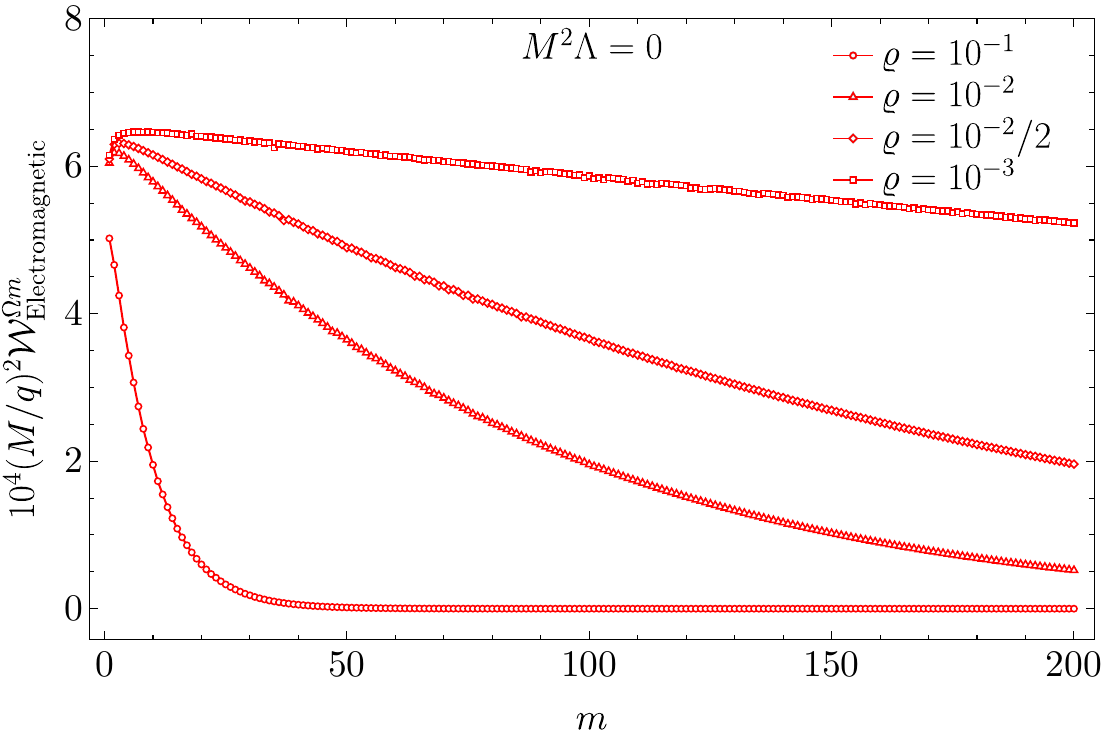}%
\includegraphics[width=0.5\linewidth]{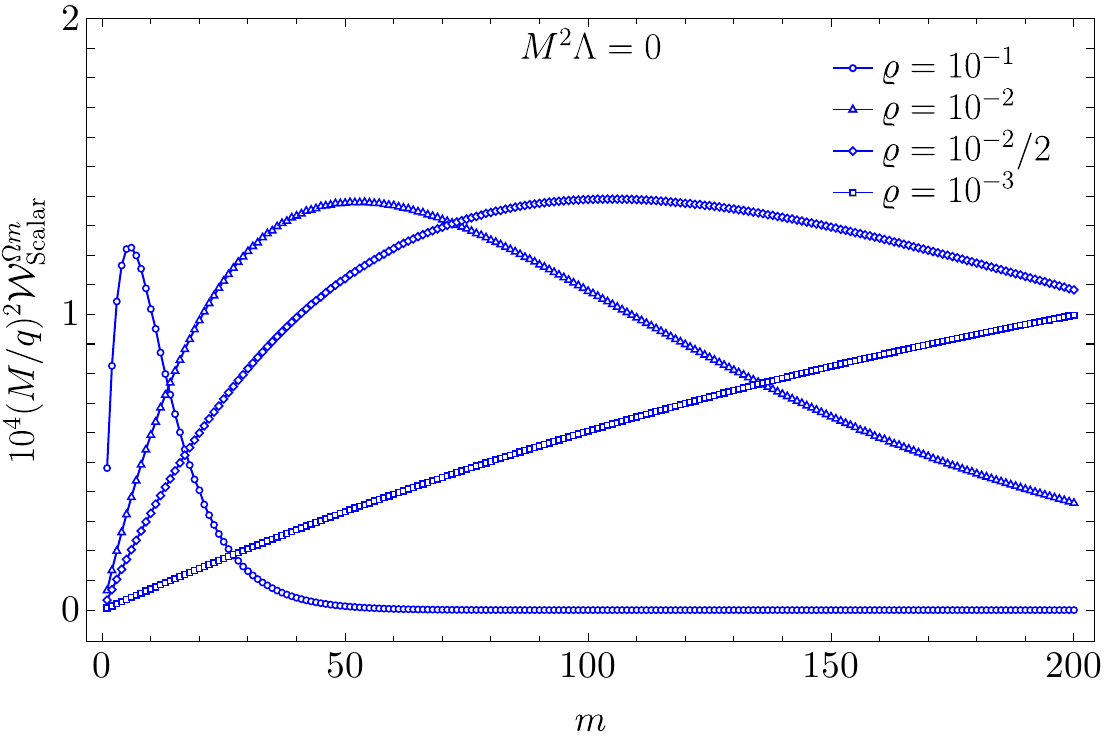}\\%
\includegraphics[width=0.5\linewidth]{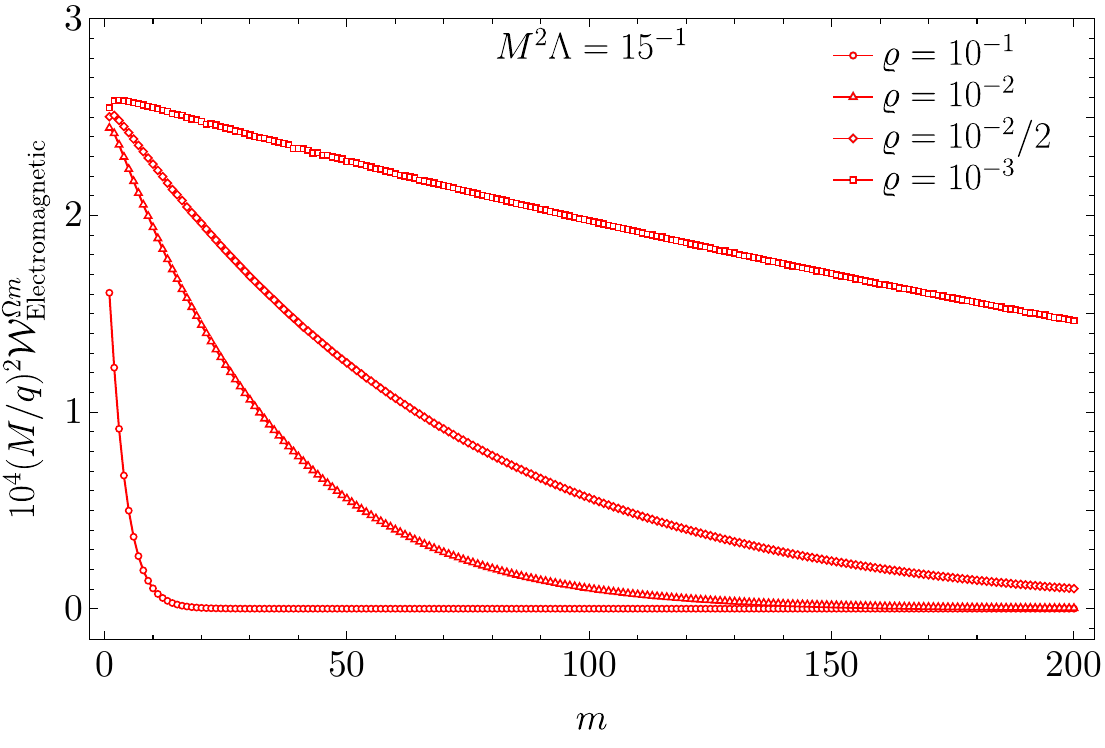}%
\includegraphics[width=0.5\linewidth]{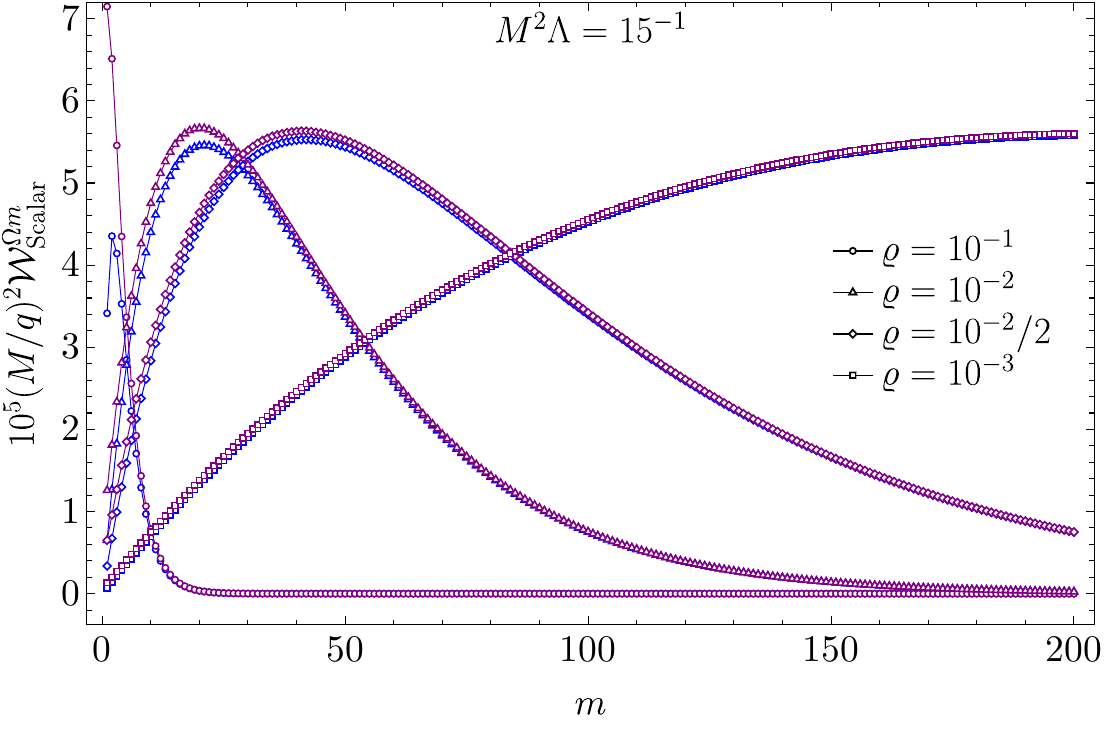}%
    \caption{Spectral distributions for electromagnetic (left) and scalar (right) fields in Schwarzschild spacetime (top) and SdS spacetime with $M^2 \Lambda = 15^{-1}$ (bottom), for different choices of $\varrho$. The bottom-right panel shows the minimal (purple) and conformal (blue) coupling cases.}
    \label{fig_spectrum_R}
\end{figure*}

 As we have seen, $\xi$ plays a role even in the strong field region in SdS spacetime, i.e., it affects the emitted power and spectral distributions of orbits near the photon sphere. In addition, here we compare the electromagnetic and scalar spectra for different orbits near the photon sphere in Schwarzschild and SdS spacetimes, i.e., for orbits located at $R=r_0+\varrho M$, with $\varrho \ll 1$. As shown in Fig.~\ref{fig_spectrum_R}, the scalar field spectra in SdS spacetime exhibit a prominent peak at a frequency $\omega_{m_0}$ greater than the fundamental frequency, $\Omega$, while the electromagnetic spectrum features a much broader frequency spectrum, with a significant contribution of lower multipoles.   This behavior is consistent with that of Schwarzschild spacetime, as can be seen in the top panels of Fig.~\ref{fig_spectrum_R}. The minimal and conformal coupling cases in SdS spacetime are shown in the bottom-right panel, where the lower multipoles of the scalar spectrum are clearly enhanced for minimal coupling when compared with conformal coupling~\cite{brito_2020,brito_2024_nonmini}.
The difference between minimal and conformal coupling in the spectra becomes more pronounced as the orbital radius increases.

\section{Final Remarks}
\label{sec_remarks}
We used the semiclassical approach of quantum field theory in curved spacetimes to analyze the electromagnetic radiation emitted by a charged particle in circular orbits around a Schwarzschild--de~Sitter BH. We computed both the total power and the spectral distribution of the radiation and compared the results with previous studies of scalar fields, minimally and nonminimally coupled, obtained in Refs.~\cite{brito_2020,brito_2024_nonmini}, as well as with the electromagnetic case in the Schwarzschild background studied in Ref.~\cite{castineiras_2005} for stable orbits. In particular, we paid attention to the influence of the coupling choice on the emitted power and spectral distributions, taking the electromagnetic case as a reference.

It is found that, for orbits with low angular velocity, the total electromagnetic power tends to a value slightly smaller than twice that of the scalar field with \textit{conformal} coupling, consistent with the behavior observed in the Schwarzschild case. In contrast, for minimal coupling, the scalar power may even exceed the electromagnetic power. As previously shown in Ref.~\cite{brito_2020}, lower multipole contributions to the scalar emitted power are enhanced under minimal coupling in the SdS spacetime. It is interesting to note that, in anti-de~Sitter spacetimes, conversely, it is the higher multipoles that are enhanced~\cite{brito_2021}.

We also analyzed the electromagnetic and scalar spectral distributions due to a particle orbiting very close to the photon sphere and in orbits with low angular velocity. In particular, we found that the normalized electromagnetic and scalar spectra in Schwarzschild spacetime resemble the electromagnetic and conformally coupled scalar spectra, respectively, in SdS spacetime.

Our results suggest that it is the scalar radiation with conformal coupling in SdS spacetimes that closely resembles the massless scalar field case in Schwarzschild spacetime, where the Ricci scalar is zero, whereas scalar radiation with minimal coupling exhibits significant differences. This can be seen as supporting the specific choice of the conformal coupling constant, $\xi=1/6$, as the natural one. Other arguments also support this choice.
For example, if we require the scalar field dynamics to be consistent with the equivalence principle, conformal coupling emerges as the more natural choice~\cite{sonego_1993}.

In Schwarzschild spacetime, for ultrarelativistic orbits, the scalar spectra exhibit a sharp peak at high frequencies, characteristic of synchrotronlike emission, while the electromagnetic spectra are significantly broader, with substantial contributions from lower multipole modes. We found that this pattern also holds in the SdS spacetime.

These results complement previous studies of scalar and electromagnetic fields and contribute to a deeper understanding of dynamical processes in asymptotically de~Sitter backgrounds. In particular, they can be used as an important tool for investigating extreme mass ratio inspirals in de~Sitter-like spacetimes.
Prospective studies may also explore the generalization of our analysis to massive fields, such as Proca and massive scalar fields, to investigate how the presence of mass and additional degrees of freedom, an additional polarization state in the case of the Proca field, affects the radiation emitted by circular orbits.
\begin{acknowledgments}
The authors thank Funda\c{c}\~ao Amaz\^onia de Amparo a Estudos e Pesquisas (FAPESPA),  Conselho Nacional de Desenvolvimento Cient\'ifico e Tecnol\'ogico (CNPq), and Coordena\c{c}\~ao de Aperfei\c{c}oamento de Pessoal de N\'{\i}vel Superior (Capes)--Finance Code 001, in Brazil, for partial financial support.
This work has further been supported by the European Horizon Europe staff exchange (SE) program HORIZON-MSCA-2021-SE-01 Grant No. NewFunFiCO-101086251.

\end{acknowledgments}



\end{document}